\documentclass[conference,a4paper]{IEEEtran}

\IEEEoverridecommandlockouts
\usepackage{cite}
\usepackage{amsmath,amssymb,amsfonts}
\usepackage{amsthm}
\usepackage{algorithmic}
\usepackage{graphicx}
\usepackage{textcomp}
\usepackage{xcolor}
\usepackage[ruled,vlined,linesnumbered]{algorithm2e}
\usepackage{float}
\def\BibTeX{{\rm B\kern-.05em{\sc i\kern-.025em b}\kern-.08em
		T\kern-.1667em\lower.7ex\hbox{E}\kern-.125emX}}

\usepackage{pgfplots}
\usepgfplotslibrary{dateplot}
\pgfplotsset{compat=1.8}
\allowdisplaybreaks

\newcommand{\be}[1]{\begin{equation}\label{#1}}
	\newcommand{\ee}{\end{equation}}

\newcommand{\bc}{\begin{center}}
	\newcommand{\ec}{\end{center}}

\newcommand{\cA}{{\cal A}}

\newcommand{\cD}{{\cal D}}

\newcommand{\cQ}{{\cal Q}}

\newcommand{\cS}{{\cal S}}
\newcommand{\cT}{{\cal T}}

\newcommand{\E}{{\sf E}}

\DeclareMathOperator*{\argmax}{arg\,max}

\newcommand{\Cref}[1]{Co\-rol\-la\-ry\,\ref{#1}}

\newtheorem{theorem}{Theorem}
\newtheorem{lemma}{Lemma}

\newtheorem{prop}{Proposition$\!$}

\newtheorem{cor}{Corollary$\!$}

\newtheorem{defi}{Definition$\!$}

\newtheorem{cl}{Claim$\!$}

\newtheorem{exam}{Example$\!$}

\newtheorem{remrk}{Remark$\!$}

\newtheorem{const}{Construction$\!$}

\newcommand{\al}[1]{\begin{align}#1\end{align}}

\newcounter{numcount}
\newcommand{\sbt}{\,\begin{picture}(-1,1)(-1,-3)\circle*{3}\end{picture}\ }

\newif\iflong
\longtrue

\newif\ifshort
\shortfalse

\newif\ifdraft

\newif\ifoldversion
\oldversiontrue

\newcommand{\longversion}[1]{
	\iflong
	#1
	\else
	\fi
}

\newcommand{\shortversion}[1]{
	\ifshort
	#1
	\else
	\fi
}

\newcommand{\bluechange}[1]{{\leavevmode\color{black}{#1}}}

\usepackage[colorinlistoftodos]{todonotes}

\usepackage[numbers,sort&compress]{natbib}

\begin{document}
    \pagestyle{plain}

\title{An Instance-Based Approach to \\ the Trace Reconstruction Problem}


\author{
\IEEEauthorblockN{Kayvon Mazooji and Ilan Shomorony}
\IEEEauthorblockA{
	University of Illinois, Urbana-Champaign\\
	mazooji2@illinois.edu,  ilans@illinois.edu}
\thanks{
The work of K.M. and I.S. was supported in part by the National Science Foundation (NSF) under grants CCF-2007597 and CCF-2046991.  Part of this paper was presented at the 58th Annual Conference on Information Sciences and Systems (CISS 2024).}

\vspace{-5mm}
}
\maketitle

\begin{abstract}
In the trace reconstruction problem, one observes the output of passing a binary string $s \in \{0,1\}^n$ through a deletion channel $T$ times and wishes to recover $s$ from the resulting $T$ ``traces.''
Most of the literature has focused on characterizing the hardness of this problem in terms of the number of traces $T$ needed for perfect reconstruction either in the worst case  or in the average case (over input sequences $s$).
In this paper, we propose an alternative, instance-based approach to the problem.
We define
the ``Levenshtein difficulty'' of a problem instance $(s,T)$ as the probability that the resulting traces do not provide enough information for correct recovery with full certainty.
One can then try to characterize, for a specific $s$, how $T$ needs to scale in order for the Levenshtein difficulty to go to zero, and seek reconstruction algorithms that match this scaling for each $s$.
\bluechange{We derive a lower bound on the Levenshtein difficulty, and prove that $T$ needs to scale exponentially fast in $n$ for the Levenshtein difficulty to approach zero for a very broad class of strings.  
For a class of binary strings with alternating long runs, we design an algorithm whose probability of reconstruction error approaches zero whenever the Levenshtein difficulty approaches zero.   For this class, we also prove that the error probability of this algorithm decays to zero at least as fast as the Levenshtein difficulty.}
\end{abstract}

\section{Introduction}

In the trace reconstruction problem, originally proposed by Levenshtein in 1997 \cite{levenshtein1997reconstruction,levenshtein2001efficientIT}, there exists a binary source string $s \in \{0, 1\}^n$ and we are given a set $\cT$ of traces of $s$, where a trace is simply a subsequence of $s$.
We are then asked to reconstruct $s$ from $\cT$.
The problem received considerable attention in the last few years \cite{
batu2004reconstructing,abroshan2019coding,magner2016fundamental,mao2018models,holenstein2008trace,de2017optimal,hartung2018trace,cheraghchi2020coded,nazarov2017trace,srinivasavaradhan2018maximum,srinivasavaradhan2019symbolwise,sabary2020reconstruction}, partially due to its applications in nanopore sequencing \cite{mao2018models}, DNA-based storage \cite{yazdi_portable_2017,organick_scaling_2017} and personalized immunogenomics \cite{safonova2019novo}.



Most of the work on the trace reconstruction problem has focused on characterizing the minimum number of traces $T$ required to guarantee perfect reconstruction of  $s$ with high probability.
Batu et al. \cite{batu2004reconstructing} studied the problem in the setting where each trace is generated by passing $s$ through a deletion channel that deletes every bit in $s$ independently with probability $p$ (and coined the name ``trace reconstruction'').  
The authors 
derived lower and upper bounds on the number of traces $T$ needed to guarantee correct reconstruction with high probability in two different problem settings:
in the \emph{worst-case} trace reconstruction problem, an algorithm is required to reconstruct \emph{any} sequence $s \in \{0,1\}^n$ with high probability; in the \emph{average-case} trace reconstruction problem, $s$ is assumed to be chosen uniformly at random from $\{0,1\}^n$ (equivalently, each symbol is independently chosen as ${\rm Bern}(1/2)$).


Since the work by Batu et al., significant development~has been made for both the worst-case and the average-case versions of the problem.  
For the average-case version, assuming $s$ is chosen uniformly at random from $\{0,1\}^n$, it is known that $T = \exp( O(\log^{1/3} n))$ traces are sufficient \cite{hartung2018trace}, while at least $\tilde{\Omega}(\log^{5/2}(n))$ traces are required~\cite{Chase2}. 
However, in many practical settings the assumption that $s$ is an i.i.d.~random string is unrealistic (e.g., DNA sequences exhibit long repeat patterns~\cite{BBT,nsg,hinge}).
On the other hand, for the worst-case version of the problem, it is known that $T = \exp(\tilde{O}(n^{1/5}))$ traces suffice \cite{Chase}, while the best known lower bound on the minimum  number of traces required is $\tilde{\Omega}(n^{3/2})$~\cite{Chase2}.
The gap between the upper and lower bounds in this case is striking.
This suggests that the landscape of problem instances is too diverse, some easy and some very hard, and requiring an algorithm to reconstruct any $s$ correctly may be too strict.


Inspired by the reconstruction condition in Levenshtein's original work \cite{levenshtein1997reconstruction,levenshtein2001efficientIT}, 
we propose an instance-based approach to the trace reconstruction problem.
The approach is based on a feasibility question: if the length of the source string $s$ is known, when does the set of traces $\cT$ contain enough information to allow the unambiguous recovery of $s$ with full certainty?  
More precisely, we say that $\cT$ is Levenshtein sufficient if $s$ is the only length-$n$ sequence that could have generated $\cT$.  The term {\it Levenshtein sufficiency} is used because reconstruction with full certainty is required in Levenshtein's original study of the problem \cite{levenshtein1997reconstruction,levenshtein2001efficientIT}. 
More random traces are generally required for  Levenshtein-sufficiency to hold compared to the commonly studied goal of reconstruction with high probability \cite{Chase, Chase2}, since the latter  does not require that the string has been reconstructed with full certainty.  
Current state of the art algorithms for worst-case trace reconstruction with high probability 
can use a metric, such as likelihood, to choose among a set of possible candidate reconstructions that are all consistent with $\cT$.
In contrast, Levenshtein-sufficiency requires that there is only one possible reconstruction from the set of traces.  

For a problem instance defined by the a pair $(s,T)$ (where $T$ is the number of traces, not the traces themselves),
a natural definition for the difficulty of $(s,T)$ is then
\al{ \label{eq:diff_intro}
\cD(s,T) = 
\Pr\left( \cT \text{ is not sufficient} \,|\, s, T \right), 
}
where $\cT$ is a set of $T$ traces of $s$ generated with deletion probability $p$.  We refer to $\cD(s,T)$ as the Levenshtein difficulty of $(s,T)$.

For example, the sequence $s = 00\dots0011\dots11$ 
(similar to the sequences used to prove the lower bound in \cite{batu2004reconstructing}) 
has a high value of $\cD(s,T)$, 
since it is unlikely that the traces will reveal the right number of zeros and ones.
Levenshtein difficulty provides an algorithm-independent measure of difficulty for reconstructing a particular source string $s$ that other frameworks do not explicitly provide, and thus gives rise to an instance-based approach to trace reconstruction.    

Intuitively,  a good reconstruction algorithm should have a small error probability for instances $(s,T)$ for which $\cD(s,T)$ is small, but should not be heavily penalized if its error probability is large on an instance with large $\cD(s,T)$.
In particular, if we consider the asymptotic regime $n \to \infty$ (considered in most studies of the trace reconstruction problem), a natural goal is to design an algorithm $\cA$ satisfying
\al{ \label{eq:goal_intro}
\Pr({\rm error}\,|\, \cA, s, T) \to 0  \text{\, whenever \,} \cD(s,T) \to 0,
}
as $n \to \infty$.
This means that, if $T$ scales with $n$ fast enough so that $\cD(s,T) \to 0$ (i.e., $T$ is scaling in a way that makes the problem feasible with high probability), then the algorithm's error probability also goes to zero.

\bluechange{We initiate the investigation of this framework for the trace reconstruction problem by deriving general necessary conditions for $\cT$ to be Levenshtein sufficient for $s$.  
These conditions are in terms of specific sequence patterns that must be preserved in the traces, and they translate into lower bounds on $\cD(s,T)$.}
\bluechange{We use these necessary conditions to prove that a broad class of strings containing consecutive repeated sequences requires $\exp(\text{poly}(n))$ traces for Levenshtein sufficiency to hold with high probability. 
We then focus on a class of strings $s$ (strings that are formed by a fixed number of alternating runs of zeros and ones) and show that, for this class, if $T$ grows faster than $\exp(c^*n)$, where $c^*$ is a positive constant that depends on the length of the runs, there exists an algorithm such that $\Pr({\rm error}\,|\, \cA, s, T) \to 0$; if $T$ grows slower than that, then $\cD(s,T) \to 1$.
Therefore, for this class of strings, this 
algorithm is optimal, in the sense that its error probability goes to zero for all cases where the difficulty tends to zero.
Moreover, we show that in those cases the 
error probability decays to zero at least as fast as the instance difficulty $\cD(s,T)$.}
\shortversion{An appendix containing omitted derivations is available in a longer version of this paper \cite{mazooji2023instance_based_trace}.} 



\section{Preliminaries}
	Strings in this paper are binary and indexed starting from~1.  
 	For a given string $x$, we let $|x|$ denote the length of $x$.
	A {\it subsequence} of $x$ is a string that can be formed by deleting elements from $x$, and a {\it supersequence} of $x$ is a string that can be formed by inserting elements into $x$. This is in comparison to a {\it substring} of $x$, which is a string that appears contiguously in $x$. 
    We let $x[i, j] = (x_i, x_{i+1}, \dots, x_j)$ be the substring of $x$ that begins at position $i$ and ends at position $j$.  For a string $a$ and positive integer $r$, we let $a^r$ be the length $r|a|$ string $aa..a$ that has $a$ repeated $r$ times.  A {\it run} in a string $s$ is a  substring in $s$ of the form $0^r$ that has a $1$ or nothing on either side, or $1^r$ that has a $0$ or nothing on either side.  For example, for the string $010011$, the runs in order are $0$, $1$, $00$, and $11$.  For a string $x$, we denote the $i$th run in $x$ by $r_i(x)$.  If $x$ is clear from context, we denote $r_i(x)$ by $r_i$. Following standard notation, for two real-valued sequences $\{a_n\}$ and $\{b_n\}$, we write $a_n \sim b_n$ if $\lim_{n \to \infty} a_n / b_n = 1$. 
    We let $\log(x)$ denote the natural logarithm of $x$.  For a matrix $A$, we denote the transpose of $A$ by $A^\dagger$.  Following standard notation, for a positive integer $M$, we let $[M]$ denote the set $\{1, \dots,  M\}$.

	Let $s = (s_1, s_2, \dots, s_n) \in \{0, 1\}^n$ be the length-$n$ string we are trying to recover. $s$ will be called the {\it source string}.  A {\it trace} of $s$ is any subsequence of $s$. 
	For our probabilistic analysis, we denote the {\it channel} that $s$ is passed through by  $\mathcal{C}$.
	In this paper, $\mathcal{C}$ is the {\it deletion channel} $\text{Del}_p$ that deletes each bit of the source string $s$ independently with probability $p$.
	In our probabilistic analysis, a trace $t$ of $s$ is the output of channel  $\mathcal{C}$ when $s$ is passed through it, i.e.,   $t \leftarrow \mathcal{C}(s)$.

    \begin{defi}
    A set of traces $\cT$ is {\it Levenshtein sufficient} for reconstructing a string $s$ if the only length-$n$ string that could have given rise to $\cT$ is $s$ itself. 
    \end{defi}

    \bluechange{
    To gain intuition about Levenshtein sufficiency, consider the case when $s = 0^n$. For a set of traces $\cT$ to be Levenshtein sufficient, $\cT$ must contain $s = 0^n$ itself since any set of all-$0$ traces that doesn't include $0^n$ could have been produced by a length $n$ source string that contains a $1$.  On the other hand, the maximum likelihood length $n$ source string is $0^n$ given any set of all-$0$ traces.  Therefore, the maximum likelihood estimator will correctly output $s = 0^n$, so long as $\cT$ is non-empty.  In summary, if Levenshtein sufficiency is required, $0^n$ is the hardest possible string to reconstruct, yet if it is only required that the maximum likelihood reconstruction is equal to the source string, $s = 0^n$ is the easiest possible string to reconstruct.
    }
    
    Given $\cT$ that is sufficient for $s$, we wish to reconstruct $s$ from $\cT$.
    A problem instance is defined by a pair $(s,T)$.
    \begin{defi}
    The Levenshtein difficulty of instance $(s,T)$ is defined as
    \al{ \label{eq:diff_intro}
    \cD(s,T) = 
    \Pr\left( \cT \text{ is not sufficient} \,|\, s, T \right).
    }
    \end{defi}
    
    An algorithm $\cA$ is {\it Levenshtein efficient} for a sequence of strings $\{s_n\}$ (where $n$ indicates the length of $s_n$) and number of traces $\{T_n\}$ 
    if as $n$ increases, the probability that $\cA$ fails to reconstruct $s_n$ approaches zero whenever the instance difficulty approaches zero; 
    i.e., if
    \al{ \label{eq:goal}
    \Pr({\rm error}\,|\, \cA, s_n, T_n) \to 0  \text{ whenever } \cD(s_n,T_n) \to 0,
    }
    as $n \to \infty$.
    Here, the probability of error for $\cA$ depends on any randomness used in  $\cA$ in addition to randomness in trace generation.   
    In principle, we can always find the set of all possible length-$n$ strings that could have given rise to $\cT$ by taking the intersection of all sets $S_i$ for  $i \in [T]$, where $S_i$ is the set of  length-$n$ strings that are supersequences of trace $t_i$.  
    This brute-force ``algorithm'' reconstructs the source string correctly whenever $\cT$ is Levenshtein sufficient, and is therefore a Levenshtein efficient reconstruction algorithm for any sequence of source strings $\{s_n\}$ and number of traces $\{T_n\}$. 
    However, it is not computationally efficient since in the worst case; it runs in time exponential in $n$. \bluechange{Furthermore, it is not clear how to analyze the probability of success for such an algorithm.} 
    



\section{Main Results}



Our first main result provides a lower bound on how $T_n$ must scale with the string length $n$ to guarantee that $D(s_n, T_n) \to 0$ when $s_n$ belongs to the class of sequences of strings defined below. 
\begin{defi}
We denote by $\cQ(r_n, f_n)$ the set of sequences of strings $\{s_n\}$ such that $s_n$ is a string of length $n$ that contains a substring of the form $A^{ f_n }$ for some string $A$ where $|A| = r_n$. Note that if for some $n$, $f_n$ is not an integer, the substring in $s_n$ can be $A^{ \lceil f_n \rceil }$ or   $A^{ \lfloor f_n \rfloor }$, so long as $|s_n| = n$.
\end{defi}

For example, the class $\cQ(2, n/4)$ contains the sequence of strings of the form $(01)^{n/4} 0^{n/2}$ where $A$ is taken to be $01$ in this case. 

\begin{theorem} \label{thm:main-1}
Let $c^* = \ell \log(\frac{1}{1-p^r})$.
For a sequence of strings $\{s_n\} \in \cQ(r,\ell n)$ where $r, \ell$ are constants such that $r \geq 1$ and $ 0 < \ell \leq 1 $,
the instance difficulty satisfies, as $n\to \infty$,
\al{
& \cD(s_n,T_n) \to 1 \text{ if } T_n = O(\exp\left(cn\right)), \;\; c < c^*.
}
\end{theorem}

This theorem shows that any string in $\cQ(r,\ell n)$ requires an exponential number of traces in $n$, and can be further generalized to strings in $\cQ(r,\ell n^a)$ for $a \leq 1$ as shown in the next section.  
Interestingly, the string pairs that are used to calculate lower bounds on trace reconstuction with high probability \cite{Chase2} use strings in $\cQ(r,\ell n)$.  Theorem \ref{thm:main-1} is proved by establishing necessary conditions for a set of traces to be Levenshtein sufficient, and calculating the probability that the conditions are satisfied.


\bluechange{Using Theorem~\ref{thm:main-1}, we design an algorithm and prove that it is Levenshtein efficient when $\{s_n\}$ belongs to the more restrictive class of sequences of strings  $\cS(M,\ell^*)$ defined below.
}


\begin{defi}
We denote by $\cS(M,\ell^*)$ the set of sequences of strings $\{s_n\}$
such that 
$s_n$ has $M$ runs for all $n$, the $i$th run has length $\ell_i n$ 
with $\sum_{i=1}^M \ell_i = 1$, and $\ell^* = \max_{i \in [M]} \ell_i$.
Note that if for some $n$, there exists a subset $G \subseteq [M]$ such that $\ell_i n$ is not an integer for $i \in G$, the length of the $i$th run can be chosen to be $\lceil \ell_i n \rceil$ or $\lfloor \ell_i n \rfloor$ for  each $i \in G$, in any way such that  $|s_n| = n$.
\end{defi}


For example, the class $\cS(3,1/2)$ contains strings of the form $0\dots01\dots10\dots0$ and $1\dots10\dots01\dots1$, with a maximum run of length $n/2$.
Notice that $\cS(M,\ell^*) \subset \cQ(1, \ell^*n).$
\bluechange{
While somewhat restrictive, this class is interesting because we can design a Levenshtein efficient algorithm with a probability of error that tends towards zero at least as fast as the Levenshtein difficulty.  
} 
\bluechange{
\begin{cor} \label{cor:long_runs}
Let $c^* = \ell^* \log(\frac{1}{1-p})$.
For a sequence of strings $\{s_n\} \in \cS(M,\ell^*)$,
the instance difficulty satisfies, as $n\to \infty$,
\al{
& \cD(s_n,T_n) \to 1\text{ if }T_n = O(\exp\left(cn\right)), \;\; c < c^*.
}
\end{cor}
}




\bluechange{Corollary \ref{cor:long_runs} follows directly from Theorem \ref{thm:main-1}.}
It may seem counterintuitive that $T=\exp(\Omega(n))$ traces are required for a set of traces to be Levenshtein-sufficient with high probability for a source string in $\cS(M,\ell^*)$, while $T=\exp(\tilde{O}(n^{1/5}))$ traces are known to be sufficient for reconstructing any source string with high probability \cite{Chase}.  This difference in trace complexity occurs because algorithms for reconstructing a string with high probability can select one among a set of possible reconstructions based on some specific criterion (such as likelihood or another statistic of the traces), without the set of traces necessarily being Levenshtein sufficient.


\bluechange{
In Theorem \ref{thm:main2}, we state that the Maximal Runs algorithm (Algorithm \ref{alg:run_matching}) performs well whenever $T_n$ scales so that 
 $\cD(s_n,T_n) \to 0$.
While very simple, the Maximal Runs algorithm is computationally efficient in that it runs in $O(nT)$ time.
}

\bluechange{
\begin{theorem} \label{thm:main2}
Let $c^* = \ell^* \log(\frac{1}{1-p})$, and let $\cA$ be the Maximal Runs algorithm (Algorithm~\ref{alg:run_matching}).
For $\{s_n\} \in \cS(M,\ell^*)$, $\cA$ satisfies, 
\al{
& \lim_{n\to \infty} \Pr({\rm error}\,|\, \cA, s_n, T_n) = 0, \\
& \liminf_{n\to \infty} \frac{\log(\Pr({\rm error}\,|\, \cA, s_n, T_n))}{\log(\cD(s_n,T_n))} \geq 1 \label{eq:logratio}}

as long as $T_n = \Omega(\exp\left(cn\right))$, for $c > c^*$.
\end{theorem}
}

\bluechange{Observe that $c^*$ is a sharp threshold on the exponent $c$ in $T = \exp(cn)$, below which the Levenshtein difficulty approaches one as $n$ increases, and above which, the Maximal Runs algorithm outputs the correct answer with high probability as $n$ increases.  The Maximal Runs algorithm is therefore Levenshtein efficient.}
\bluechange{Furthermore, (\ref{eq:logratio}) implies that whenever $\cD(s,T)$ goes to zero, the error probability $\Pr({\rm error}\,|\, \cA, s_n, T_n)$ goes to zero at an exponential rate that is at least as fast as the rate with which $\cD(s,T)$ goes to zero.
}


\begin{algorithm}
		\caption{Maximal Runs}\label{alg:run_matching}
		\KwData{$n, \cT$}	\KwResult{$\hat{s}$}
            $\hat{s} \gets $ empty string\;
		$\hat{M} \gets$ maximum number of runs in any trace in $\cT$\;
		$S \gets$ set of all traces with $\hat{M}$ runs\;
		\If{$t_1[1] = t_2[1]$  for all $ t_1, t_2 \in S$} {
		\For{$i \in [\hat{M}]$}
		{
            $t^* \gets \arg \max_{t \in S} |r_i(t)|$ \;  
            $x_i \gets r_i(t^*)$\; }
		\If{$\sum_{i \in [\hat{M}]} |x_i| = n $}
		{
		$\hat{s} \gets x_1 x_2 ... x_{\hat{M}}$\;
		}
		}
\end{algorithm}


\section{Concluding remarks}

In this paper, we proposed an instance-based approach to the trace reconstruction problem based on a new notion of instance-specific difficulty.\bluechange{
For a class of strings with a fixed number of runs, we derived an exponential bound on the number of traces, below which, the instance difficulty to goes to one, and above which, a simple algorithm reconstructs the string with high probability.
While the class of strings considered is somewhat restrictive, we prove any number of traces above the bound leads to perfect reconstruction with high probability, in contrast to most existing results for trace reconstruction.}
In addition, we derived a lower bound on the number of traces for the instance difficulty to go to zero for a much broader class of strings. 
This work can thus be seen as developing the initial tools for a more general characterization of the instance-based hardness of trace reconstruction.
We note that our Theorem \ref{thm:main-1} can be generalized to the following as proved in the following section.  
\begin{theorem} \label{thm:main-1_more_general}
Let $c^* = \ell \log(\frac{1}{1-p^r})$.
For a sequence of strings $\{s_n\} \in \cQ(r,\ell n^a)$ 
 where $r \geq 1$ and $ 0 < \ell, a \leq 1 $,
the instance difficulty satisfies, as $n\to \infty$,
\al{
& \cD(s_n,T_n) \to 1 \text{ if } T_n = O(\exp\left(cn^a\right)), \;\; c < c^*
}
\end{theorem}
This shows that any string that contains a constant length string repeated consecutively a polynomial number of times in $n$ requires a superpolynomial number of traces in $n$ for Levenshtein sufficiency to hold with high probability.

Finally, we point out that other interesting definitions for the ``sufficiency'' of $\cT$ are possible.
For example, one could say $\cT$ is sufficient for $s$ if the maximum likelihood source string given $\cT$ is $s$.  

\section{Proof of Main Results}


We begin by introducing necessary 
conditions for  $\cT$ to be Levenshtein sufficient for $s$, which we will use to prove the main results.
Out of the conditions in Lemma~\ref{lem:nec}, we only need condition (1)  for our analysis of Levenshtein sufficiency, but we include the other two conditions  to show additional requirements for Levenshtein sufficiency.

\begin{lemma} \label{lem:nec}
	For any two distinct binary strings $A, B$ such that $|A| \leq |B|,$ we have the following necessary conditions on the set of traces $\cT$ to be Levenshtein sufficient for $s$.  Let $a,b \in \mathbb{N}$ such that $a,b \geq 1.$
	\begin{enumerate}
    	\item If $A^a$ is a substring of $s,$ it does not happen that for every trace in $\cT$, there exists a copy of $A$ in this substring that is deleted.
    	\item If $B^a A B^b$ is a substring of $s,$ it does not happen that for every trace in $\cT$, there exists a copy of $B$ in this substring that is deleted.
    	\item  If $A^a B^b$ or $B^b A^a$ is a substring of $s,$ it does not happen that for every trace in $\cT$, there exists a copy of $A$ or $B$ that is deleted from this substring.
	\end{enumerate}
\end{lemma}

\begin{IEEEproof}
        (1) If a copy of $A$ is deleted from this substring in every trace, then $\cT$ could arise from $s$ with the substring $A^a$ replaced by $DA^{a-1}$ for any string $D$ such that $|D| = |A|$ and $D \neq A$.
        
        (2) If a copy of $B$ is deleted from this substring in every trace, then $\cT$ could arise from $s$ with the substring $B^a A B^b$ replaced by $G = B^{a-1} A B A 1^{|B|-|A|} B^{b-1}$.  To see this, notice that if a trace of $s$ has a copy of $B$ deleted from $B^a$ in $B^a A B^b$, then the same trace can be formed if $B^a A B^b$ is replaced by $G$ in $s$ since $A 1^{|B|-|A|)}$ can be deleted from $G$.  A similar argument holds if a trace of $s$ has a copy of $B$ deleted from $B^b$ in $B^a A B^b$.
        
        (3) If a copy of $A$ or $B$ is deleted from the substring  $A^a B^b$ in every trace, then $\cT$ could arise from $s$ with the substring $A^a B^b$ replaced by $A^{a-1} BA B^{b-1}$.  The $B^b A^a$ case is analogous.
\end{IEEEproof}

\bluechange{
\begin{lemma} \label{lem:suff}
    Algorithm \ref{alg:run_matching} reconstructs $s$ correctly if $\cT$ is such that for each run $i$, there exists a trace such that no run is fully deleted and run $i$ is fully preserved.
\end{lemma}
}
\bluechange{
It is easy to see that the conditions in Lemma \ref{lem:suff} guarantee that Algorithm~\ref{alg:run_matching} recovers $s$ correctly.
Also notice that the sufficient condition in Lemma \ref{lem:suff} is very similar to the necessary condition (1) in Lemma \ref{lem:nec} when $A$ is a single bit, and this observation forms the basis of our result.  
}


We now prove the main theorems.  
For a source string $s_n$ where $\{s_n\} \in \cQ(r, f_n)$ and $r$ is a constant, let $E_1^{x_n}$ denote the event that necessary condition (1) holds for the substring $x_n$ in $s_n$ where $x_n = A^{f_n}$ for some string $A$ of length $r$.
In other words, $E_1^{x_n}$ is the event that for the substring $x_n=AA...A$ of interest, we have that there exists a trace where no copy of $A$ in $x_n$ is fully deleted.  
\bluechange{Let $E_2$ denote the event that the reconstruction conditions for Algorithm~\ref{alg:run_matching} in Lemma \ref{lem:suff} holds for $(s_n,T_n)$.}


\bluechange{
Notice that for any $s_n $ and $T_n$ such that $\{s_n\} \in  \cQ(r, f_n)$ and $x_n$ is a substring of $s_n$ of the form described above, it follows that
\al{ \label{eq:d_bounds}
\Pr(\bar E_1^{x_n}) \leq \cD(s_n,T_n).
}
Therefore, by proving that  $\lim_{n \to \infty} \Pr(\bar{E}_1^{x_n}) = 1$ for a pair of sequences $\{s_n\}, \{T_n\}$, we prove that $\lim_{n \to \infty} \cD(s_n,T'_n) = 1$ for any $\{T'_n\}$ such that $T'_n = O(T_n)$ since $\cD(s,T)$ can only increase for fixed $s$ if $T$ decreases.}
\bluechange{
Lemma \ref{lem:main} gives an asymptotic characterization of $\Pr(\bar E_1^{x_n})$ for $\{s_n\}  \in  \cQ(r, \ell n^a)$ which immediately yields Theorems \ref{thm:main-1} and \ref{thm:main-1_more_general}. 
Lemma~\ref{lem:main2} gives an asymptotic characterization of $\Pr(\bar E_2)$.  The first statement in Lemma \ref{lem:main} applied to $\cS(M,\ell^*)$ (which is a subset of $\cQ(1, \ell^* n)$), along with Lemma~\ref{lem:main2}, immediately yields Theorem \ref{thm:main2}.}

\begin{lemma} \label{lem:main}
    Suppose $\{s_n\} \in \cQ(r, \ell n^a)$ where $r, \ell, a$ are constants such that  $0<\ell, a \leq 1$, and let $x_n$ be a substring of $s_n$ of the form $A^{\ell n^a}$ where $|A| = r$.
    Let $c^* = \ell \log(\frac{1}{1-p^r})$. Then, as $n \to \infty$,  
    \al{
    & \sbt \;\; \Pr(\bar{E}^{x_n}_1) \to 0 \text{ if }T_n = \Omega(\exp\left(cn^a\right)), \;\; c > c^* \label{eq:limcabove} \\
    & \sbt \;\; \Pr(\bar{E}^{x_n}_1) \to 1/e \text{ if }T_n = \exp\left(c^*n^a\right) \label{eq:limcequal} \\
    & \sbt \;\; \Pr(\bar{E}^{x_n}_1) \to 1\text{ if }T_n = O(\exp\left(cn^a\right)), \;\; c < c^* \label{eq:limcbelow} 
    }
\end{lemma}


\begin{lemma} \label{lem:main2}
    Suppose $\{s_n\} \in \cS(M,\ell^*)$ and let $x_n$ be a run in $s_n$ of length $\ell^* n$.  
    Let $c^* = \ell^* \log(\frac{1}{1-p})$. Then, as $n \to \infty$,  
    \al{
    & \frac{\log(\Pr(\bar{E}_2))}{\log(\Pr(\bar{E}^{x_n}_1))} \to 1\text{ if }T_n = \Theta(\exp\left(cn\right)), \;\; c > c^*.  \label{eq:limratio}
    }
\end{lemma}


\section{Proof of Lemmas~\ref{lem:main} and \ref{lem:main2}}
    For ease of presentation in this proof, we write $T_n$ as $T$.  For the source string $s_n,$ let $x_n$ be a substring of $s_n$ of the form $A^{f_n}$ where $A$ is of length $r$.  
    We have that 
    \begin{align}
		\Pr(E_1^{x_n}) & = 1 - \Pr(\bar{E}_1^{x_n}) = 1 - (1 - (1-p^{r})^{f_n})^T.  \nonumber
    \end{align}

    Let $E_2^i$ be the event that there is at least one trace that has the $i$th run fully preserved and has no run fully deleted.  With slight abuse of notation, let $r_i$ be the length of the $i$th run.  Let $a$ be the $T \times 1$ binary vector that has a $1$ in the $i$th position if the $i$th trace has no run fully deleted, and has a $0$ in the $i$th position otherwise.
	We then have that
	\begin{align}
		\Pr(E_2) & = \sum_{a \in \{0,1\}^{T \times 1}} \Pr(E_2 | a) \Pr(a) \nonumber
		\\ & = \sum_{a} \left(\prod_{i=1}^M \Pr(E_2^i | a)\right) \left(\prod_{i=1}^M (1-p^{r_i})\right)^{1^\dagger a} \nonumber 
		\\ & \quad \times \left(1 - \prod_{i=1}^M (1-p^{r_i})\right)^{T - 1^\dagger a} \nonumber
		\\ & = \sum_{a} \left(\prod_{i=1}^M \left(1 - \left(1 - \frac{(1-p)^{r_i}}{1 - p^{r_i}} \right)^{1^\dagger a}\right) \right) \nonumber
		\\ & \quad \times \left(\prod_{i=1}^M (1-p^{r_i})\right)^{1^\dagger a} \left(1 - \prod_{i=1}^M (1-p^{r_i})\right)^{T - 1^\dagger a} \nonumber
		\\ & = \sum_{j = 0}^T {T \choose j} \left(\prod_{i=1}^M \left(1 - \left(1 - \frac{(1-p)^{r_i}}{1 - p^{r_i}} \right)^{j}\right) \right) \nonumber
		\\ & \quad \times (\prod_{i=1}^M (1-p^{r_i}))^{j} (1 - \prod_{i=1}^M (1-p^{r_i}))^{T - j}
	\end{align}
	where the third equality follows because 
	\begin{align}
		& \Pr(E_2^i | a) = 1 - \Pr(\bar{E}_2^i | a) \nonumber
		\\ & = 1 - \frac{\Pr(\bar{E}_2^i \cap a)}{\Pr(a)} \nonumber
		\\ & = 1 - \frac{ (1-p^{r_i} - (1-p)^{r_i})^{1^\dagger a} }{\left(\prod_{j=1}^M (1-p^{r_j})\right)^{1^\dagger a} \left(1 - \prod_{j=1}^M (1-p^{r_j})\right)^{T - 1^\dagger a}} \nonumber
		\\ & \quad \times \left(\prod_{j \neq i}^M (1-p^{r_j})\right)^{1^\dagger a} \left(1 - \prod_{j=1}^M \left(1-p^{r_j}\right)\right)^{T - 1^\dagger a}  \nonumber
		\\ & = 1 - \frac{ (1-p^{r_i} - (1-p)^{r_i})^{1^\dagger a}}{( 1-p^{r_i})^{1^\dagger a} } \nonumber
		\\ & = 1 - \left(1 - \frac{ (1-p)^{r_i}}{ 1-p^{r_i} }\right)^{1^\dagger a}.
	\end{align}
	Observe that  
	\begin{align}
		\Pr(E_2) = \E \left[\prod_{i=1}^M \left(1 - \left(1 - \frac{(1-p)^{r_i}}{1 - p^{r_i}} \right)^{X}\right) \right]
	\end{align}
	where $X \sim \text{Bin}(T, \; \prod_{i=1}^M (1-p^{r_i}))$ is a binomial random variable with $T$ trials and probability parameter $\prod_{i=1}^M (1-p^{r_i})$.

	\subsection{Analysis of $\Pr(\bar{E}^x_1)$}

    Suppose the string $x_n = AA...A$ that we are analyzing is such that $A$ is repeated $\ell n^a$ times where $0 < a, \ell \leq 1$ are constants, and $|A| = r.$ where $r>0$ is constant.

    Suppose the number of traces is $T = \exp(cn^a)$ where $c$ is a positive  constant. We will write $n$ in terms of $T$ in the expression for $\Pr(\bar{E}^{x_n}_1)$ to perform asymptotic analysis. 
    According to the formula in the previous section, and letting $q = \ell / c$, we have 
    \begin{align}
	\Pr(\bar{E}^{x_n}_1) &  =  (1-(1-p^r)^{\ell n^a})^T 
  \nonumber 
        \\ & 
        =  (1-(1-p^r)^{q \log(T)})^T \nonumber
        \\ & \sim \exp \left(    -T^{q \log(1-p^r) + 1} \right). \label{eq:notE1_asymptotic}
    \end{align}
      \shortversion{We prove the asymptotic expression in (\ref{eq:notE1_asymptotic}) in the appendix \cite{mazooji2023instance_based_trace}.}
      \longversion{We prove the asymptotic expression in (\ref{eq:notE1_asymptotic}) in the appendix.}      
    
    If $c > c^* = \ell \log(\frac{1}{1-p^r})$, which is equivalent to $\frac{\ell}{c} \log(1-p^r) + 1 > 0 \quad \forall i \in [M]$, 
    then 
    \begin{align}
        \lim_{n \to \infty} \Pr(\bar{E}^{x_n}_1) = 0.
    \end{align}
	If $T$ grows faster than $\exp(c^*n^a)$, i.e., $T = \Omega(\exp(cn^a))$ for $c > c^*$,
	then $P(\bar{E}^{x_n}_1)$ approaches zero because 
for fixed $n$, having more traces can only cause $P(\bar{E}^{x_n}_1)$ to decrease.
This proves (\ref{eq:limcabove}).
On the other hand, if $c 
 < c^*$, then clearly 
 \begin{align}		    \lim_{n \to \infty} \Pr(\bar{E}^{x_n}_1) = 1.
\end{align}
If $T =  O(\exp(cn^a))$ where $c < c^*$, we have that $\lim_{n \to \infty} \Pr(\bar{E}^{x_n}_1) = 1$ since for any such $c$, there exists a larger value of $c$ that also satisfies the property and for fixed $n$, having less traces can only cause $P(\bar{E}^{x_n}_1)$ to increase.
This proves  (\ref{eq:limcbelow}).
Finally, if $c 
 = c^*$, then clearly 
 \begin{align}		    \lim_{n \to \infty} \Pr(\bar{E}^{x_n}_1) = 1/e.
\end{align}

	\subsection{Analysis of $\Pr(\bar{E}_2)$}

In this section suppose that $\{s_n\} \in \cS(M, \ell^*)$.
Notice that $\cS(M, \ell^*) \subset \cQ(1, \ell^* n)$, so the analysis of $\Pr(\bar{E}_1^{x_n})$ in the previous subsection can be applied to any string in $\cS(M, \ell^*)$.
 
	Suppose the number of traces is $T_n = \exp(cn)$ where $c$ is a positive  constant.
	For ease of presentation, let $q_i = \ell_i/c$ and $u_i = \frac{\ell_i}{c} \log(T)$. Let $X \sim \text{Bin}(T, \; \prod_{i=1}^M (1-p^{u_i}))$ be the binomial random variable with $T$ trials and probability parameter $p_X = \prod_{i=1}^M (1-p^{u_i})$. We have that
	\begin{align}
		& \Pr(\bar{E}_2) = 1 - \E \left[ \prod_{i=1}^M \left(1 - \left(1 - \frac{(1-p)^{u_i}}{1 - p^{u_i}} \right)^{X}\right) \right] \nonumber
		\\ & = 1 - \E \left[ 1 - \sum_{y = 1}^M \; \sum_{\substack{K \subseteq [M] : \\ |K| = y}} (-1)^{y+1} \prod_{i \in K}  \left(1 - \frac{(1-p)^{u_i}}{1 - p^{u_i}} \right)^{X} \right] \nonumber
		\\ & = \sum_{y = 1}^M \; \sum_{K \subseteq [M] : \; |K| = y} (-1)^{y+1} \E \left[ \prod_{i \in K}  \left(1 - \frac{(1-p)^{u_i}}{1 - p^{u_i}} \right)^{X} \right] \nonumber
		\\ & = \sum_{y = 1}^M \; \sum_{K \subseteq [M] : \; |K| = y} (-1)^{y+1} \nonumber
		\\ & \quad \times \E \left[ \exp \left(\log \left(\prod_{i \in K}  \left(1 - \frac{(1-p)^{u_i}}{1 - p^{u_i}} \right)\right) X\right) \right] \nonumber
		\\ & = \sum_{y = 1}^M \! \!\!\! \!\!\!\!\!\!\!\!\!\!\!\!\!\sum_{\quad \quad \quad K \subseteq [M] : \; |K| = y} \!\!\!\!\!\!\!\!\!\!\!\!\!\!\!\!\!\!\! (-1)^{y+1} \!\!\left( \! 1 \!- p_X \!+ p_X \prod_{i \in K}\left(1 - \frac{(1-p)^{u_i}}{1 - p^{u_i}} \right) \right)^T
	\end{align}
	from the moment-generating function of a binomial random variable.
Letting $N = |\{i \; : \; \ell_i = \ell_{i^*} \}|$ where $i^* = \argmax_i \ell_i$, we have that  $\Pr(\bar{E}_2)$ is asymptotically given by 
	\begin{align}
	    \!\!\!\! N \exp \left( -   \left( \prod_{k=1}^M (1-T^{q_k \log(p)}) \right) \left( \frac{T^{q_{i^*}\log(1-p) + 1}}{1 - T^{q_{i^*} \log(p)}}  \right) \right) \label{eq:notE2_asymptotic}
	\end{align}
\shortversion{as proved in the appendix \cite{mazooji2023instance_based_trace}.}
\longversion{as proved in the appendix.}
Therefore, if $c > \ell_{i^*}   \log(\frac{1}{1-p}) = \ell^*\log(\frac{1}{1-p}) = c^*$,
	\begin{align}
	    & \frac{\log(\Pr(\bar{E}_2))}{\log(\Pr(\bar{E}^{r_{i^*}(s_n)}_1))}  
\sim \frac{\prod_{k=1}^M (1-T^{q_k \log(p)})}{1 - T^{q_{i^*} \log(p)}} \sim 1 \label{eq:logratio2}
	\end{align}
\shortversion{as proved in the appendix \cite{mazooji2023instance_based_trace}.}
\longversion{as proved in the appendix.}
 This concludes the proof of Lemma~\ref{lem:main2}. 
{\footnotesize
{
\bibliographystyle{ieeetr}
\bibliography{refs.bib}

\begin{thebibliography}{10}

\bibitem{levenshtein1997reconstruction}
V.~Levenshtein, ``Reconstruction of objects from a minimum number of distorted patterns,'' {\em Doklady Mathematics}, vol.~55, no.~3, pp.~417--420, 1997.

\bibitem{levenshtein2001efficientIT}
V.~I. Levenshtein, ``Efficient reconstruction of sequences,'' {\em IEEE Transactions on Information Theory}, vol.~47, no.~1, pp.~2--22, 2001.

\bibitem{batu2004reconstructing}
T.~Batu, S.~Kannan, S.~Khanna, and A.~McGregor, ``Reconstructing strings from random traces,'' {\em Departmental Papers (CIS)}, p.~173, 2004.

\bibitem{abroshan2019coding}
M.~Abroshan, R.~Venkataramanan, L.~Dolecek, and A.~G. i~F{\`a}bregas, ``Coding for deletion channels with multiple traces,'' in {\em 2019 IEEE International Symposium on Information Theory (ISIT)}, pp.~1372--1376, IEEE, 2019.

\bibitem{magner2016fundamental}
A.~Magner, J.~Duda, W.~Szpankowski, and A.~Grama, ``Fundamental bounds for sequence reconstruction from nanopore sequencers,'' {\em IEEE Transactions on Molecular, Biological and Multi-Scale Communications}, vol.~2, no.~1, pp.~92--106, 2016.

\bibitem{mao2018models}
W.~Mao, S.~N. Diggavi, and S.~Kannan, ``Models and information-theoretic bounds for nanopore sequencing,'' {\em IEEE Transactions on Information Theory}, vol.~64, no.~4, pp.~3216--3236, 2018.

\bibitem{holenstein2008trace}
T.~Holenstein, M.~Mitzenmacher, R.~Panigrahy, and U.~Wieder, ``Trace reconstruction with constant deletion probability and related results.,'' in {\em SODA}, vol.~8, pp.~389--398, 2008.

\bibitem{de2017optimal}
A.~De, R.~O'Donnell, and R.~A. Servedio, ``Optimal mean-based algorithms for trace reconstruction,'' in {\em Proceedings of the 49th Annual ACM SIGACT Symposium on Theory of Computing}, pp.~1047--1056, 2017.

\bibitem{hartung2018trace}
L.~Hartung, N.~Holden, and Y.~Peres, ``Trace reconstruction with varying deletion probabilities,'' in {\em 2018 Proceedings of the Fifteenth Workshop on Analytic Algorithmics and Combinatorics (ANALCO)}, pp.~54--61, SIAM, 2018.

\bibitem{cheraghchi2020coded}
M.~Cheraghchi, R.~Gabrys, O.~Milenkovic, and J.~Ribeiro, ``Coded trace reconstruction,'' {\em IEEE Transactions on Information Theory}, 2020.

\bibitem{nazarov2017trace}
F.~Nazarov and Y.~Peres, ``Trace reconstruction with exp (o(n1/3)) samples,'' in {\em Proceedings of the 49th Annual ACM SIGACT Symposium on Theory of Computing}, pp.~1042--1046, 2017.

\bibitem{srinivasavaradhan2018maximum}
S.~R. Srinivasavaradhan, M.~Du, S.~Diggavi, and C.~Fragouli, ``On maximum likelihood reconstruction over multiple deletion channels,'' in {\em 2018 IEEE International Symposium on Information Theory (ISIT)}, pp.~436--440, IEEE, 2018.

\bibitem{srinivasavaradhan2019symbolwise}
S.~R. Srinivasavaradhan, M.~Du, S.~Diggavi, and C.~Fragouli, ``Symbolwise map for multiple deletion channels,'' in {\em 2019 IEEE International Symposium on Information Theory (ISIT)}, pp.~181--185, IEEE, 2019.

\bibitem{sabary2020reconstruction}
O.~Sabary, A.~Yucovich, G.~Shapira, and E.~Yaakobi, ``Reconstruction algorithms for dna-storage systems,'' {\em bioRxiv}, 2020.

\bibitem{yazdi_portable_2017}
H.~T. Yazdi, R.~Gabrys, and O.~Milenkovic, ``Portable and error-free {{DNA}}-based data storage,'' {\em Scientific Reports}, vol.~7, no.~1, 2017.

\bibitem{organick_scaling_2017}
L.~Organick, S.~D. Ang, Y.-J. Chen, R.~Lopez, S.~Yekhanin, K.~Makarychev, M.~Z. Racz, G.~Kamath, P.~Gopalan, B.~Nguyen, and C.~N. Takahashi, ``Random access in large-scale dna data storage,'' {\em Nature Biotechnology}, 2018.

\bibitem{safonova2019novo}
Y.~Safonova and P.~A. Pevzner, ``De novo inference of diversity genes and analysis of non-canonical v (dd) j recombination in immunoglobulins,'' {\em Frontiers in immunology}, vol.~10, p.~987, 2019.

\bibitem{Chase2}
Z.~Chase, ``New lower bounds for trace reconstruction,'' in {\em Annales de l'Institut Henri Poincar{\'e}, Probabilit{\'e}s et Statistiques}, vol.~57, pp.~627--643, Institut Henri Poincar{\'e}, 2021.

\bibitem{BBT}
G.~Bresler, M.~Bresler, and D.~Tse, ``Optimal assembly for high throughput shotgun sequencing,'' {\em BMC Bioinformatics}, 2013.

\bibitem{nsg}
I.~Shomorony, T.~Courtade, S.~Kim, and D.~Tse, ``Information-optimal genome assembly via sparse read-overlap graphs,'' {\em Bioinformatics}, vol.~37, no.~17, 2016.

\bibitem{hinge}
G.~M. Kamath, I.~Shomorony, F.~Xia, T.~A. Courtade, and N.~T. David, ``Hinge: long-read assembly achieves optimal repeat resolution,'' {\em Genome research}, vol.~27, no.~5, pp.~747--756, 2017.

\bibitem{Chase}
Z.~Chase, ``Separating words and trace reconstruction,'' in {\em Proceedings of the 53rd Annual ACM SIGACT Symposium on Theory of Computing}, pp.~21--31, 2021.

\end{thebibliography}
}
}

\longversion{\section{Appendix}
\subsection{Proof of (\ref{eq:notE1_asymptotic})}
We have that 	
 \begin{align}
		 \Pr(\bar{E}_1^x) & = (1-(1-p^r)^{q \log(T)})^T \nonumber
  \\ & = \exp \left( T \log   (1-(1-p^r)^{q \log(T)}) \right) \nonumber
  		\\ & = \exp \left( T \log   (1-T^{q \log(1-p^r)}) \right) \nonumber
		\\ & \sim \exp \left(   T    (-T^{q \log(1-p^r)}) \right) \nonumber
		\\ & = \exp \left(    -T^{q \log(1-p^r) + 1} \right) \nonumber,
	\end{align}
where we used the fact that $\log(1-x) \sim -x$ if $x \to 0$.

\subsection{Proof of (\ref{eq:notE2_asymptotic})}
We have that $\Pr(\bar{E}_2)$ equals
	\begin{align}
		\sum_{y = 1}^M \! \!\!\! \!\!\!\!\!\!\!\!\!\!\!\!\!\sum_{\quad \quad \quad K \subseteq [M] : \; |K| = y} \!\!\!\!\!\!\!\!\!\!\!\!\!\!\!\!\!\!\! (-1)^{y+1} \!\!\left( \! 1 \!- p_X \!+ p_X \prod_{i \in K}\left(1 - \frac{(1-p)^{u_i}}{1 - p^{u_i}} \right) \right)^T. \nonumber 
	\end{align}
	Thus, we have to find an asymptotic formula for 
	\[  \left( 1 - p_X + p_X \prod_{i \in K}\left(1 - \frac{(1-p)^{u_i}}{1 - p^{u_i}} \right) \right)^T \]
	for general $K \subseteq [M].$  We have that
	\begin{align}
	    & \left( 1 - p_X + p_X \prod_{i \in K}\left(1 - \frac{(1-p)^{u_i}}{1 - p^{u_i}} \right) \right)^T \nonumber
	    \\ & = \exp\left( T\log \left( 1 - p_X + p_X \prod_{i \in K}\left(1 - \frac{(1-p)^{q_i \log(T)}}{1 - p^{q_i \log(T)}} \right) \right) \right) \nonumber
	    \\ & = \exp\left( T\log \left( 1 - p_X\left(1 - \prod_{i \in K}\left(1 - \frac{T^{q_i \log(1-p)}}{1 - T^{q_i \log(p)}} \right) \right) \right) \right) \nonumber
	    \\ & \sim \exp\left( - T  p_X\left(1 - \prod_{i \in K}\left(1 - \frac{T^{q_i \log(1-p)}}{1 - T^{q_i \log(p)}} \right) \right) \right)  \nonumber
	    \\ & = \exp\left( - T  p_X\left( \sum_{y = 1}^{|K|} \; \sum_{\substack{S \subseteq K :\\ |S| = y}} (-1)^{y+1} \prod_{i \in S} \frac{T^{q_i \log(1-p)}}{1 - T^{q_i \log(p)}} \right) \right)  \nonumber
	    \\ & = \exp\left( -  p_X\left( \sum_{y = 1}^{|K|} \; 
	    \sum_{\substack{S \subseteq K :\\ |S| = y}} (-1)^{y+1} \prod_{i \in S} \frac{T^{q_i \log(1-p)+ 1/|S|}}{1 - T^{q_i \log(p)}} \right) \right)  \nonumber
	\end{align}
	Observe that if $q_i \log(1-p) + 1 > 0$ for all $i \in [M]$ (i.e. $c > \max_i \ell_i \log(\frac{1}{1-p})$), any term for a non-singleton $K = Z$ is little-o of the term for $K = \{z\}$ where $z \in Z$.   Thus, only the terms corresponding to singleton $K$ in $\Pr(\bar{E}_2)$ can matter asymptotically, leaving us with the following asymptotic formula for $\Pr(\bar{E}_2)$:
	\begin{align}
		\sum_{i=1}^M \exp \left( -   \left( \prod_{k=1}^M (1-T^{q_k \log(p)}) \right) \left( \frac{T^{q_i \log(1-p) + 1}}{1 - T^{q_i \log(p)}}  \right) \right) \nonumber
	\end{align}
	Out of these $M$ terms, only the ones corresponding to $i^* = \argmax_i \ell_i$ matter asymptotically.  Letting $N = |\{i \; : \; \ell_i = \ell_{i^*} \}|$, we have that  $\Pr(\bar{E}_2)$ is asymptotically given by 
	\begin{align}
	    N \exp \left( -   \left( \prod_{k=1}^M (1-T^{q_k \log(p)}) \right) \left( \frac{T^{q_{i^*}\log(1-p) + 1}}{1 - T^{q_{i^*} \log(p)}}  \right) \right). \nonumber
	\end{align}

\subsection{Proof of (\ref{eq:logratio2})}
Letting $N = |\{i \; : \; \ell_i = \ell_{i^*} \}|$ where $i^* = \argmax_i \ell_i$, we have that 
	if $c > \ell_{i^*}   \log(\frac{1}{1-p}) = \ell^*\log(\frac{1}{1-p}) = c^*$,

 
	\begin{align}
	    & \frac{\log(\Pr(\bar{E}_2))}{\log(\Pr(\bar{E}^{r_{i^*}(s_n)}_1))}  \nonumber
		\\ & \sim \frac{ \log\left(N \exp \left( -   \left( \prod_{k=1}^M (1-T^{q_k \log(p)}) \right) \left( \frac{T^{1 + \log(1-p)  q_{i^*}}}{1 - T^{q_{i^*} \log(p)}}  \right) \right)\right)}{ \log\left( \exp \left( - T^{1 + \log(1-p)  q_{i^*}
		}  \right) \right)} \nonumber
		\\ & = \frac{\log(1/N) +   \left( \prod_{k=1}^M (1-T^{q_k \log(p)}) \right) \left( \frac{T^{1 + \log(1-p)  q_{i^*}}}{1 - T^{q_{i^*} \log(p)}} \right)}{ T^{1 + \log(1-p)  q_{i^*}}} \nonumber
		\\ & \sim \frac{\prod_{k=1}^M (1-T^{q_k \log(p)})}{1 - T^{q_{i^*} \log(p)}} \sim 1.
        \nonumber
	\end{align}

}

\end{document}